\documentclass[prl,twocolumn,showpacs,floatfix,amsfonts]{revtex4}
\usepackage{graphicx,graphics,color,epsfig}
\usepackage{bm}
\usepackage{amsmath}
\usepackage{amssymb}
\begin{document}
\preprint{}
\title{Quantum Electronic Transport through a Precessing Spin}
\author{Jian-Xin Zhu}
\affiliation{Theoretical Division, MS B262, Los Alamos National
Laboratory, Los Alamos, New Mexico 87545}
\author{A. V. Balatsky}
\affiliation{Theoretical Division, MS B262, Los Alamos National
Laboratory, Los Alamos, New Mexico 87545}

\begin{abstract}
{The conductance through a local nuclear spin precessing in a
magnetic field is studied by using the equations-of-motion
approach. The characteristics of the conductance is determined by
the tunneling matrix and the position of equilibrium chemical
potential. We find that the spin flip coupling between the
electrons on the spin site and the leads produces the conductance
oscillation.  When the spin is precessing in the magnetic field at
Larmor frequency ($\omega_{L}$), the conductance develops the
oscillation with the frequency of both $\omega_{L}$ and
2$\omega_{L}$ components, the relative spectrum weight of which
can be tuned by the chemical potential and the spin flip coupling.
}
\end{abstract}
\maketitle

There has been intensified interest in the electronic transport
through atomic impurities or quantum dots in condensed matter
physics. The novel features arising from the quantization of both
the electronic spectrum and the electronic charge on these
impurities have been well studied. More recently, the behavior of
a single magnetic spin has also received much attention. The
single spin detection and manipulation will play a major role in
spintronics and quantum information processing. In spintronics,
spins can be used as elementary information storage
units~\cite{Prinz98,Wolf01}. In the realm of quantum
computing~\cite{Kane98,Loss98}, several architecture proposals
rely crucially on the ability to manipulate and detect single
spins. So far, the possibility of a single spin observation is a
challenging issue. The standard electron spin detection technique
- electron spin resonance (ESR) - is limited to a macroscopic
number of electron spins - $10^{10}$ or more~\cite{Farle98}. The
state-of-the-art magnetic resonance force microscopy has recently
achieved the resolution of about 100 fully polarized electron
spins~\cite{Bruland98}. The atomic resolution of the scanning
tunneling microscope (STM) can provide an alternative technique
for the single spin detection. Manassen {\em et al.}~\cite{Mana00}
carried out the STM measurement of the tunneling current while
scanning the surface of Si in the vicinity of a local spin
impurity (Fe cluster) or imperfection (oxygen vacancy in Si-O) in
an external magnetic field. More recently, a similar STM
experiment was also performed on organic molecules by Durkan and
Welland~\cite{Durkan02}. Both experiments detected a small signal
in the current power at the Larmor frequency. The extreme
localization of the signal around the spin site prompted the
authors to attribute the detected signal to the Larmor precession
of the single spin site. Motivated by Ref.~\cite{Mana00}, it has
been proposed~\cite{Balats02} that the spin-orbit interaction of
the conduction electrons in the two-dimensional surface may couple
the tunneling current to the precessing spin. Instead, the authors
of Ref.~\cite{Mozy01} argued that it is not the spin impurity
itself but the current itself that generates the small ac signal.
The mechanism for the observed phenomenon is an issue of current
debate. In this paper, we address a new spintronic system in which
a magnetic spin is weakly coupled to leads. The low-temperature
transport through this magnetic spin is presented by using
equations of motion approach. In addition to the interest in its
own right,  the obtained picture can shed new lights on the
mechanism for the above experimental observation.

\begin{figure}
\centerline{\psfig{figure=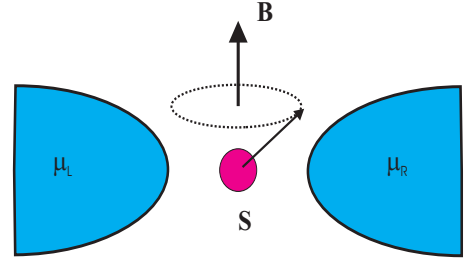,height=3.5cm,width=6cm,angle=0}}
\caption{Magnetic spin coupled to two leads. In the presence of a
magnetic field ${\bf B}$, the spin precesses around the field
direction.} \label{FIG:SETUP}
\end{figure}

The model system under consideration is illustrated in
Fig.~\ref{FIG:SETUP}. It consists of two ideal leads coupled to a
single magnetic spin. In the presence of a magnetic field, the
spin precesses around the field direction. We assume that there
are no electron-electron interaction and spin-orbit coupling
within the metallic leads. The spin-orbit interaction is confined
in both barriers between the leads and the single spin site. This
experimental setup is different from that studied in
Ref.~\cite{Mozy01}, where the single spin in Fig.~\ref{FIG:SETUP}
is replaced by a quantum dot with the spin-degenerate two levels
directly split by the magnetic field. We model the spin and its
leads by the Hamiltonian:
\begin{eqnarray}
\mathcal{H}&=&\sum_{k \in (L,R),\sigma} \epsilon_{k\sigma}
c_{k\sigma}^{\dagger}c_{k\sigma} + J\sum_{\sigma,\sigma^{\prime}}
d_{\sigma}^{\dagger}\Omega_{\sigma\sigma^{\prime}}d_{\sigma^{\prime}}
\nonumber \\
&&+ \sum_{k\in (L,R),\sigma;\sigma^{\prime}}
(V_{k\sigma,\sigma^{\prime}}c_{k\sigma}^{\dagger}d_{\sigma^{\prime}}
+\mbox{H.c.})\;. \label{EQ:Hamil}
\end{eqnarray}
Here $c_{k\sigma}^{\dagger}$ ($c_{k\sigma}$) creates (destroys) an
electron with momentum $k$ and spin $\sigma$ in either the left
($L$) or the right ($R$) lead, and $d_{\sigma}^{\dagger}$
($d_{\sigma}$) is the creation (annihilation) operator of the
single electron with spin $\sigma$ at the magnetic spin site. The
quantity  $\epsilon_{k\sigma}$ are the single particle energies of
conduction electrons in the two leads. The electrons on the spin
impurity site is connected to those in the two leads with the
tunneling matrix elements $V_{k\sigma,\sigma^{\prime}}$. The
single electron on the impurity site is coupled to the local spin
through a direct exchange interaction of strength $J$. The form of
the coupling matrix $\hat{\Omega}$ in Eq.~(\ref{EQ:Hamil}) will be
discussed below. Since the Zeeman coupling of the electrons on the
impurity site to the external magnetic field is usually very small
compared with the exchange coupling to the local spin, this
interaction term has been neglected.

The local magnetic spin $\mathbf{S}$ is defined in a
three-dimensional spin space. In an external magnetic field
$\mathbf{B}$, a torque will act on the magnetic moment $\bm{\mu}$
of amount $\bm{\mu}\times \mathbf{B}$, where
$\bm{\mu}=\gamma\bm{S}$ with $\gamma$ the gyromagnetic ratio. The
equation of motion of the local spin is given by
$\frac{d\bm{\mu}}{dt}=\bm{\mu} \times (\gamma \mathbf{B})$. For a
static magnetic field applied along the $z$ direction, we shall
see that the local spin would precess about the field in the
absence of friction. The coupling matrix then becomes:
\begin{equation}
\hat{\Omega}=\left( \begin{array}{cc} \Omega_{\uparrow\uparrow} &
\Omega_{\uparrow\downarrow} \\
\Omega_{\downarrow\uparrow}& \Omega_{\downarrow\downarrow}
\end{array}\right)=
\left( \begin{array}{cc} \cos\theta &
\sin\theta e^{-i\phi} \\
\sin\theta e^{i\phi} & \cos\theta
\end{array}\right)\;.
\label{EQ:SPIN}
\end{equation}
In Eq.~(\ref{EQ:SPIN}), $\theta$ is the angle between $\bm{\mu}$
and $\mathbf{B}$, and $\phi=-\omega_{L}t+\phi_{0}$ where
$\omega_{L}=\gamma B$ is the Larmor frequency and $\phi_{0}$ is
the initial azimuthal angle. If the friction is present between
the spin and its environment, the local spin would eventually
become parallel to the field. The friction corresponds to the
relaxation processes characterized by time $T$~\cite{Note1}.
Therefore, we assume that $T$ is sufficiently long for the
precession to be well defined. In the end of the paper, we will
make a remark for the case of a finite $T$.

Since the energy associated with the spin precession,
$\hbar\omega_{L}\sim 10^{-6}\;\mbox{eV}$ is much smaller than the
typical electronic energy on the order of 1 eV, the spin
precession is very slow as compared to the time scale of all
conduction electron process. This fact allows us to treat the
electronic problem adiabatically as if the local spin is static
for every instantaneous spin orientation~\cite{Balats02}. Our aim
is to calculate the conductance through the spin impurity. In a
generalized B\"{u}ttiker-Landauer formalism~\cite{Butt86}, it can
be expressed as~\cite{Meir92}:
\begin{equation}
g=\frac{e^{2}}{h}\int d\epsilon f^{\prime}_{\mbox{\tiny
FD}}(\epsilon) \mbox{Im}[\mbox{Tr}\{
\frac{2\hat{\Gamma}_{L}\hat{\Gamma}_{R}}{\hat{\Gamma}_{L}
+\hat{\Gamma}_{R}} \hat{G}^{r}(\epsilon)\}]\;. \label{EQ:COND}
\end{equation}
Equation~(\ref{EQ:COND}) expresses the linear-response conductance
in terms of the transmission probability weighted by the
derivative of the Fermi distribution function, $f_{\mbox{\tiny
FD}}=1/\{\exp[(\epsilon-\mu)/k_{B}T]+1\}$, with $\mu$ the chemical
potential in the equilibrium state. The transmission probability
is constructed as a product of the elastic coupling to the leads
and the Green function of electrons on the spin impurity site. The
coupling to the leads is represented by the full line-width
function:
\begin{equation}
\Gamma^{L(R)}_{\sigma\sigma^{\prime}}=2\pi
\sum_{k,\sigma^{\prime\prime} \in L(R)}
V_{k,\sigma^{\prime\prime};\sigma}^{*}V_{k,\sigma^{\prime\prime};\sigma^{\prime}}
\delta (\epsilon-\epsilon_{k})\;. \label{EQ:LINEWIDTH}
\end{equation}
Here we have assumed that the couplings to the left and right
leads are factorizable, i.e., $\hat{\Gamma^{L}}=\lambda
\hat{\Gamma^{R}}$~\cite{Meir92}. The quantity in
Eq.~(\ref{EQ:COND}) $\hat{G}^{r}(\epsilon)$ is the Fourier
transform of the retarded Green function for electrons on the spin
impurity site: \begin{equation}
G_{\sigma\sigma^{\prime}}=-i\Theta(t)
\langle[d_{\sigma}(t),d^{\dagger}_{\sigma^{\prime}}(0)]_{+}\rangle
\;,
\end{equation}
where $[\dots]_{+}$ denotes the anticommutator, $\Theta(t)$ is the
Heaviside step function, and $d_{\sigma}^{\dagger}(t)$
($d_{\sigma}(t)$) are the impurity-site electron operators in the
Heisenberg picture, e.g.,
$d_{\sigma}(t)=e^{iHt}d_{\sigma}e^{-iHt}$. Note that both
$\hat{\Gamma}$ and $\hat{G}^{r}$ are matrices in the spin space of
the impurity-site electron.

\begin{figure}
\centerline{\psfig{figure=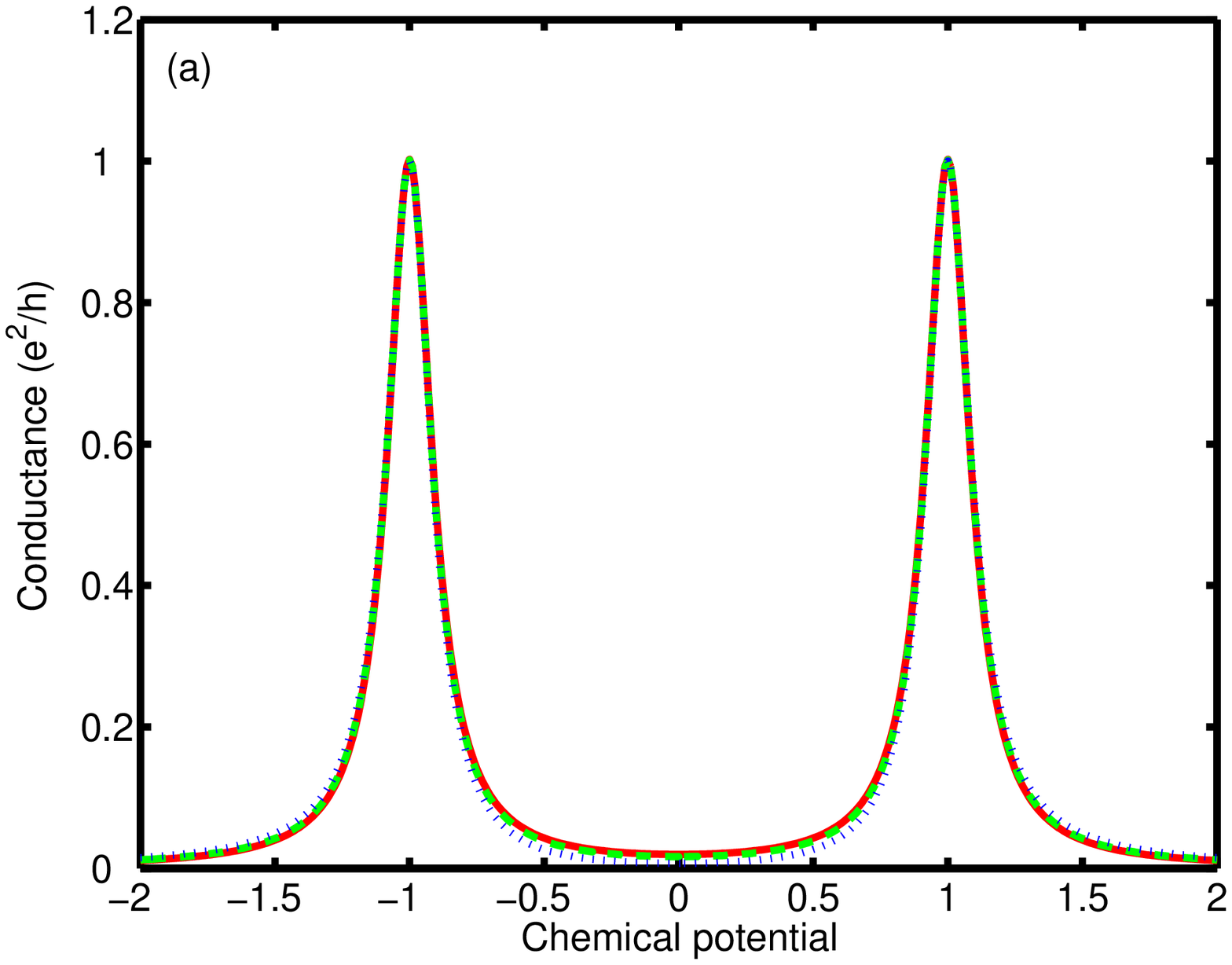,height=4.2cm,width=6cm,angle=0}}
\centerline{\psfig{figure=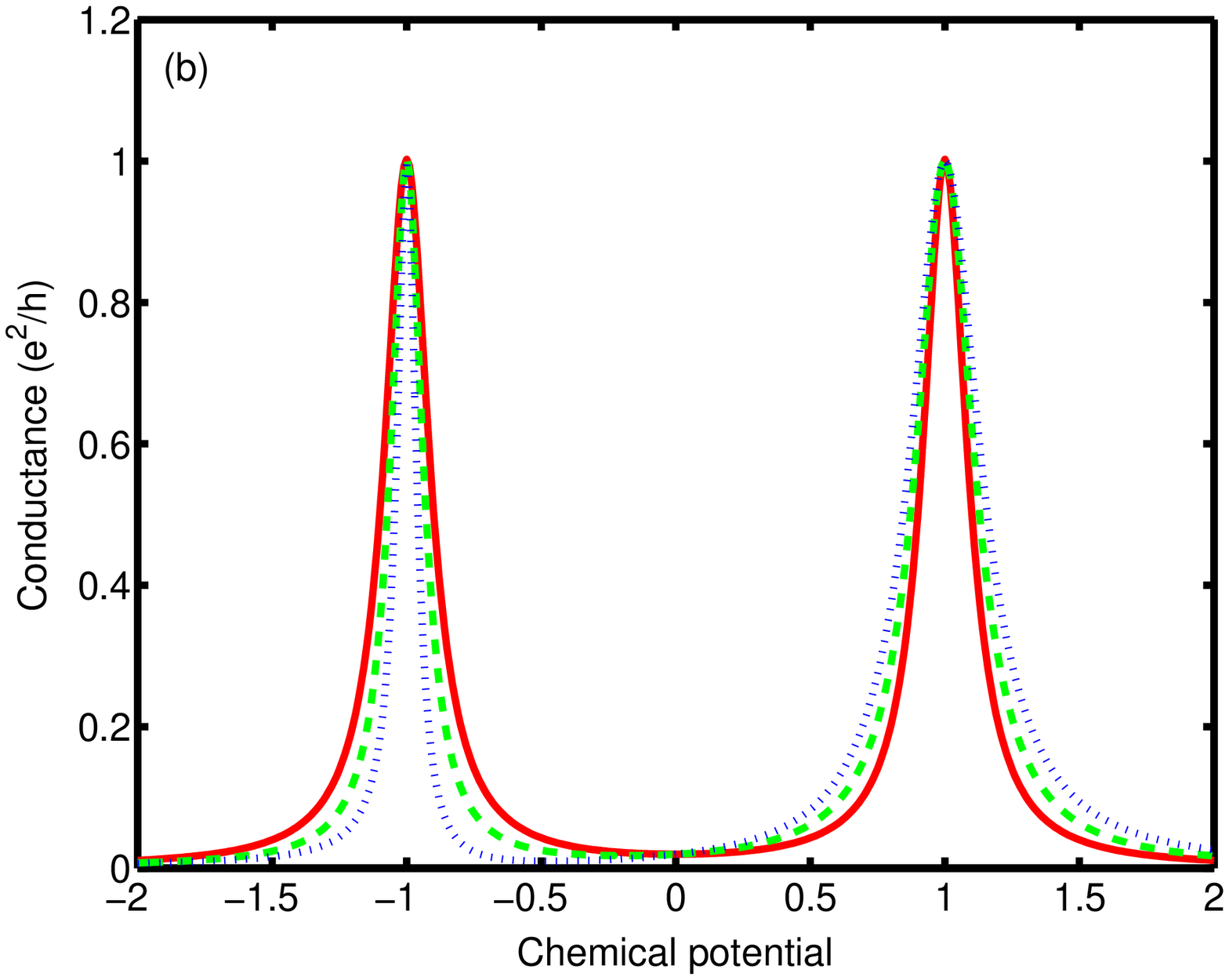,height=4.2cm,width=6cm,angle=0}}
\centerline{\psfig{figure=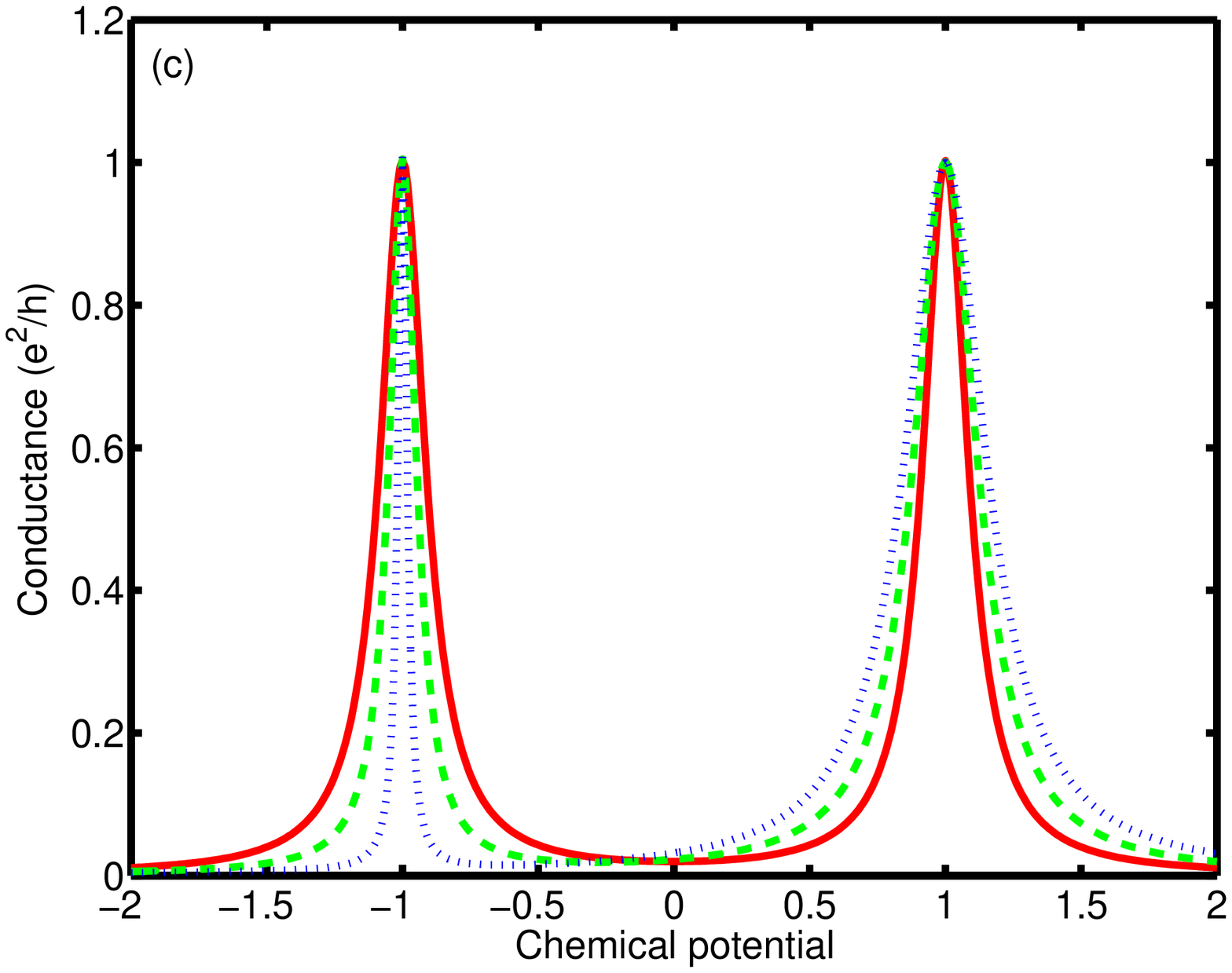,height=4.2cm,width=6cm,angle=0}}
\caption{Conductance versus the chemical potential $\mu$ with
various values of the spin-flip coupling
$\Gamma_{s}/\Gamma_{n}=0.0$ (red-solid line), 0.4 (green-dashed
line), and 0.8 (blue-dotted line). The results shown in panels (a)
through (c) correspond to three different local spin orientations
in zero magnetic field: ($\theta,\phi_0$)=($0,0$), ($\pi/4,0$),
($\pi/2,0$). The spin-conserved coupling, $\Gamma_{n}=0.1$.}
\label{FIG:COND1}
\end{figure}

The remaining task is to evaluate the impurity-site electron Green
function. This calculation should be done in the presence of the
coupling to leads. We employ the equations-of-motion method to
this end. It consists of differentiating
$G_{\sigma\sigma^{\prime}}$ with respect to time, thereby
generating new Green functions. As we will show below, in the
absence of electron correlation, the equations of motion for
$G_{\sigma\sigma^{\prime}}$ can be closed exactly. Otherwise, the
higher-order equations-of-motion arising from electron correlation
must be truncated to close the equations-of-motion for
$G_{\sigma\sigma^{\prime}}$. Using the commutator $[d_{\sigma},
H]_{-}$, we find the equation of motion for
$G_{\sigma\sigma^{\prime}}$:
\begin{eqnarray} i\frac{\partial
G_{\sigma\sigma^{\prime}}(t)}{\partial
t}&=&\delta(t)\delta_{\sigma\sigma^{\prime}}
+J\sum_{\sigma^{\prime\prime}}
\Omega_{\sigma\sigma^{\prime\prime}}G_{\sigma^{\prime\prime}\sigma^{\prime}}(t)
\nonumber \\
&&+\sum_{k,\sigma^{\prime\prime}\in L(R)}
V_{k,\sigma^{\prime\prime};\sigma}^{*}
G_{k\sigma^{\prime\prime},\sigma^{\prime}}(t)\;. \label{EQ:GREEN1}
\end{eqnarray}
The time derivative of $G_{\sigma\sigma^{\prime}}$ generates a new
Green function:
\begin{equation} G_{k\sigma,\sigma^{\prime}}
=-i\Theta(t)
\langle[c_{k\sigma}(t),d^{\dagger}_{\sigma^{\prime}}(0)]_{+}\rangle\;,
\end{equation}
which is originated from the coupling of the impurity-site
electron to the leads. Using the commutator $[c_{k\sigma},H]_{-}$,
we can also obtain: \begin{equation} i\frac{\partial
G_{k\sigma,\sigma^{\prime}}(t)}{\partial
t}=\epsilon_{k}G_{k\sigma,\sigma^{\prime}}(t)
+\sum_{\sigma^{\prime\prime}}
V_{k\sigma,\sigma^{\prime\prime}}G_{\sigma^{\prime\prime}\sigma^{\prime}}(t)\;.
\label{EQ:GREEN2}
\end{equation}
Equation for $G_{k\sigma,\sigma^{\prime}}$~(\ref{EQ:GREEN1}) now
closes with the aid of Eq.~(\ref{EQ:GREEN2}). Performing the
Fourier transform of these Green functions, we obtain
Eqs.~(\ref{EQ:GREEN1}) and (\ref{EQ:GREEN2}) in the frequency
space: \begin{eqnarray} \omega G_{\sigma\sigma^{\prime}}(\omega)
&=&\delta_{\sigma\sigma^{\prime}}+J\sum_{\sigma^{\prime\prime}}
\Omega_{\sigma\sigma^{\prime\prime}}G_{\sigma^{\prime\prime}\sigma^{\prime}}(\omega)
\nonumber \\
&&+\sum_{k,\sigma^{\prime\prime}\in L(R)}
V_{k,\sigma^{\prime\prime};\sigma}^{*}
G_{k\sigma^{\prime\prime},\sigma^{\prime}}(\omega)\;,
\label{EQ:GREEN3}
\end{eqnarray}
and
\begin{equation} \omega G_{k\sigma,\sigma^{\prime}}(\omega)
=\epsilon_{k}G_{k\sigma,\sigma^{\prime}}(\omega)
+\sum_{\sigma^{\prime\prime}}
V_{k\sigma,\sigma^{\prime\prime}}G_{\sigma^{\prime\prime}\sigma^{\prime}}(\omega)\;.
\label{EQ:GREEN4}
\end{equation}
A little algebra yields the solution: \begin{subequations}
\begin{eqnarray}
G_{++}(\omega)&=&\frac{1}{\omega-(J\Omega_{++}+\Sigma_{++})
-\Sigma_{1}} \;,
\\
G_{+-}(\omega)&=&\frac{(J\Omega_{+-}+\Sigma_{+-})G_{--}(\omega)}
{\omega-(J\Omega_{++}+\Sigma_{++})}\;,
 \\
G_{-+}(\omega)&=&\frac{(J\Omega_{-+}+\Sigma_{-+})G_{++}(\omega)}
{\omega-(J\Omega_{--}+\Sigma_{--})}\;,
\\
G_{--}(\omega)&=&\frac{1}{\omega-(J\Omega_{--}+\Sigma_{--})-\Sigma_{2}}\;.
\end{eqnarray}
\label{EQ:GREEN}
\end{subequations}
In Eq.~(\ref{EQ:GREEN}), the self-energy matrix due to the
coupling to the leads is:
\begin{equation}
\Sigma_{\sigma\sigma^{\prime}}(\omega)=\sum_{k,\sigma^{\prime\prime}\in
L,R}
\frac{V_{k\sigma^{\prime\prime},\sigma}^{*}V_{k\sigma^{\prime\prime},\sigma^{\prime}}
}{\omega-\epsilon_{k}}\;, \label{EQ:SELF}
\end{equation}
and
\begin{eqnarray}
\Sigma_{1}(\omega)&=&\frac{(J\Omega_{+-}+\Sigma_{+-})(J\Omega_{-+}+\Sigma_{-+})}
{\omega-(J\Omega_{--}+\Sigma_{--})}\;,\\
\Sigma_{2}(\omega)&=&\frac{(J\Omega_{-+}+\Sigma_{-+})(J\Omega_{+-}+\Sigma_{+-})}
{\omega-(J\Omega_{++}+\Sigma_{++})}\;.
\end{eqnarray}
Equation~(\ref{EQ:SELF}) shows that the structure of the self
energy is determined by the nature of the tunneling matrix
elements $V_{k\sigma,\sigma^{\prime}}$. The retarded self energy
can be written in terms of the principle and imaginary parts:
\begin{eqnarray}
\Sigma_{\sigma\sigma^{\prime}}^{r}(\omega)&=& \mathcal{P}
\sum_{k,\sigma^{\prime\prime}\in L,R}
\frac{V_{k\sigma^{\prime\prime},\sigma}^{*}V_{k\sigma^{\prime\prime},\sigma^{\prime}}
}{\omega-\epsilon_{k}}\nonumber \\
&&-\frac{i}{2}[\Gamma_{\sigma\sigma^{\prime}}^{L}(\omega)
+\Gamma_{\sigma\sigma^{\prime}}^{R}(\omega)]\;,
\label{EQ:FULLSELF}
\end{eqnarray}
where the full line-width functions have been given by
Eq.~(\ref{EQ:LINEWIDTH}). The solution for $\hat{G}^r(\omega)$, as
given by Eq.~(\ref{EQ:GREEN}), can now be employed to evaluate the
conductance by Eq.~(\ref{EQ:COND}). The structure of the full
line-width functions , $\hat{\Gamma}^{L,R}$, is determined by the
details of the tunneling matrix $V_{k\sigma,\sigma^{\prime}}$. We
use the full line-width functions as the coupling parameters. The
full retarded self-energies $\hat{\Sigma}^{r}$ are evaluated
according to Eq.~(\ref{EQ:FULLSELF}). For simplicity, we assume
symmetric tunneling barriers between the local spin and the leads,
therefore, $\hat{\Gamma}^{L}=\hat{\Gamma}^{R}=\hat{\Gamma}$. Since
the density of states around the Fermi surface in the leads are
broad and flat, the couplings are constant. In the following
analysis, we measure the energy in units of the exchange integral
between the electrons and the local spin, $J$, and consider the
conductance at zero temperature.  In zero magnetic field, the spin
is static. In Fig.~\ref{FIG:COND1}, we plot the conductance versus
the chemical potential with various values of the spin-flip
coupling $\Gamma_{+-}=\Gamma_{-+}=\Gamma_{s}$. The results shown
in panels (a)-(c) correspond to the three different spin
orientations: ($\theta,\phi_0$)=($0,0$), ($\pi/4,0$), ($\pi/2,0$).
The spin-conserved couplings are taken to be
$\Gamma_{++}=\Gamma_{--}=\Gamma_{n}=0.1$. From
Fig.~\ref{FIG:COND1}, the generic feature of the conductance as a
function of the chemical potential is: As $\mu$ passes through
$\pm 1$, the conductance exhibits the resonant behavior by showing
a peak at these energy position. The resonant level position is
independent of the local-spin orientation. The spin-flip couplings
does not change the resonant level position either. However, they
change the line-width of the resonant peaks. Explicitly, with the
increasing spin-flip couplings, the resonant peak at $-1$ is
narrowing while that at $+1$ is broadening, but both with the
maximum peak intensity almost unchanged.

\begin{figure}
\centerline{\psfig{figure=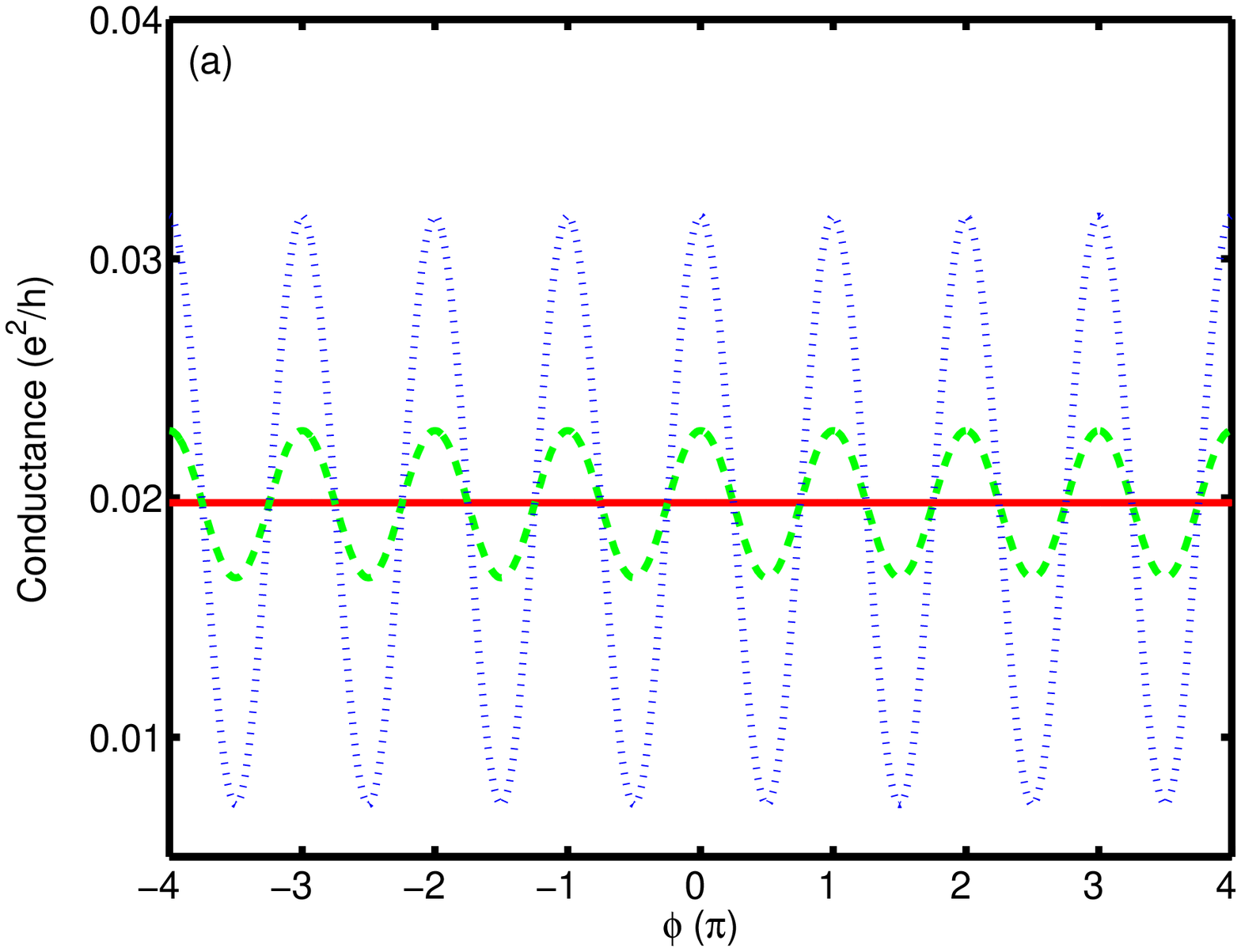,height=2.8cm,width=4cm,angle=0}
\psfig{figure=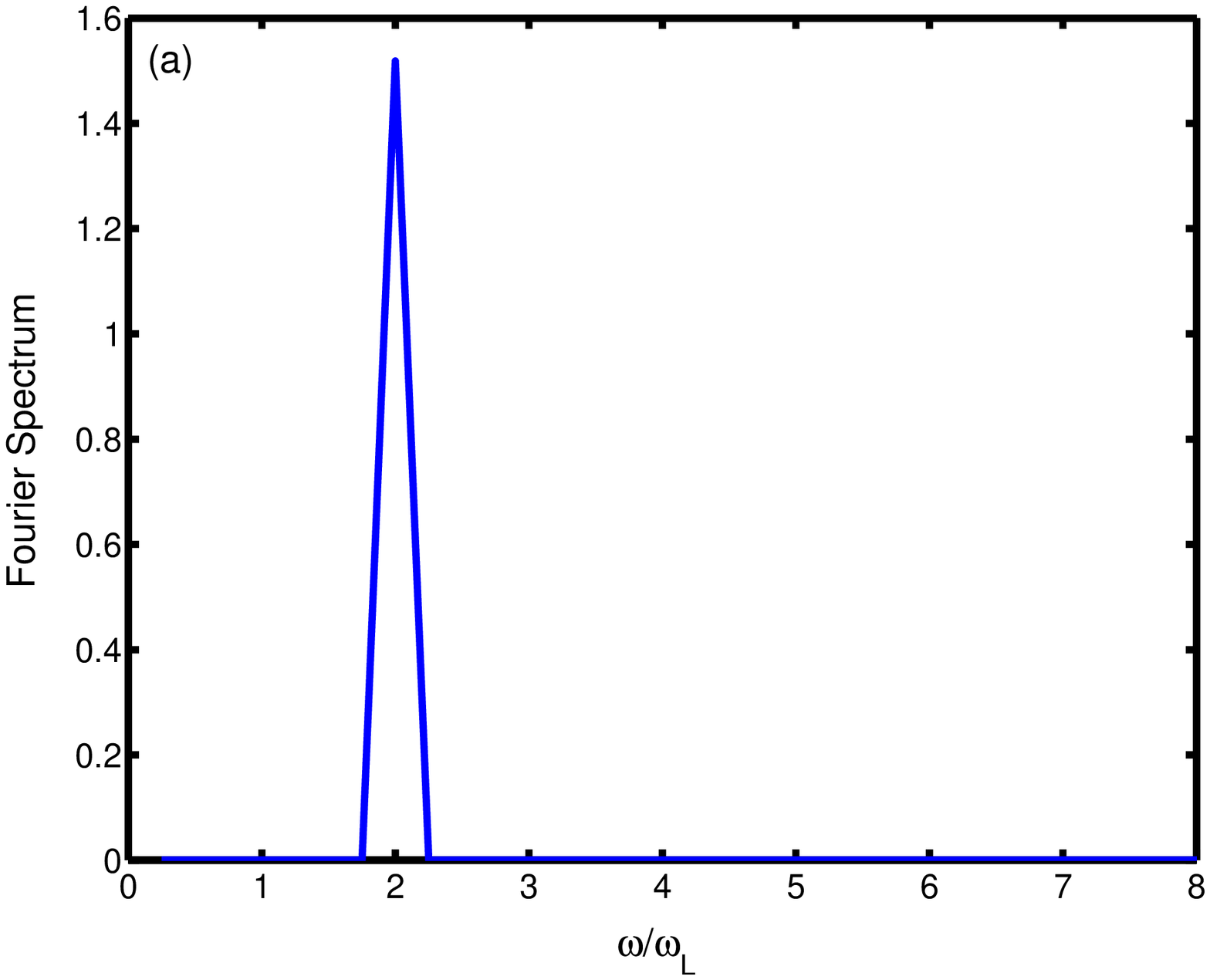,height=2.8cm,width=4cm,angle=0}}
\centerline{\psfig{figure=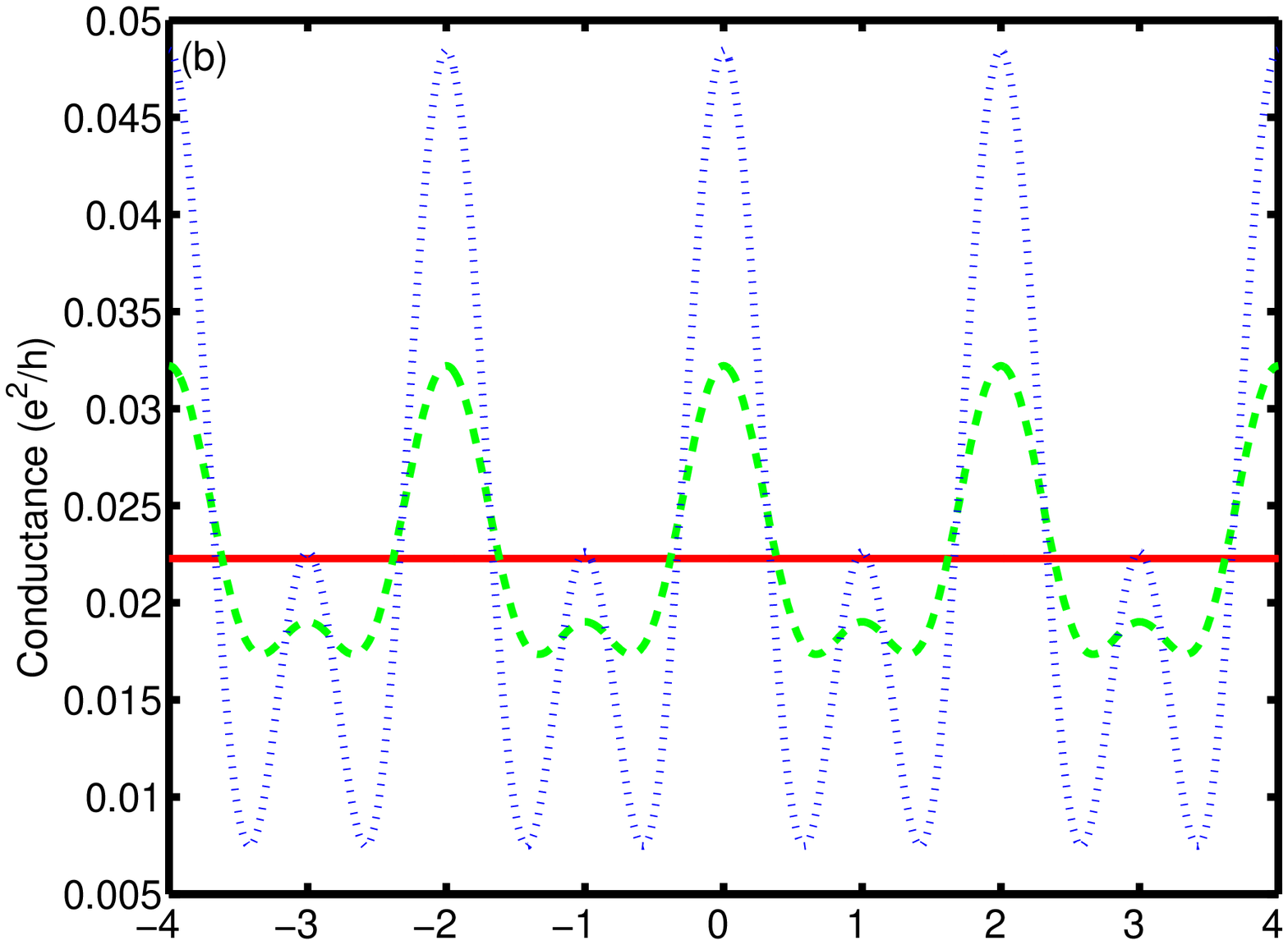,height=2.8cm,width=4cm,angle=0}
\psfig{figure=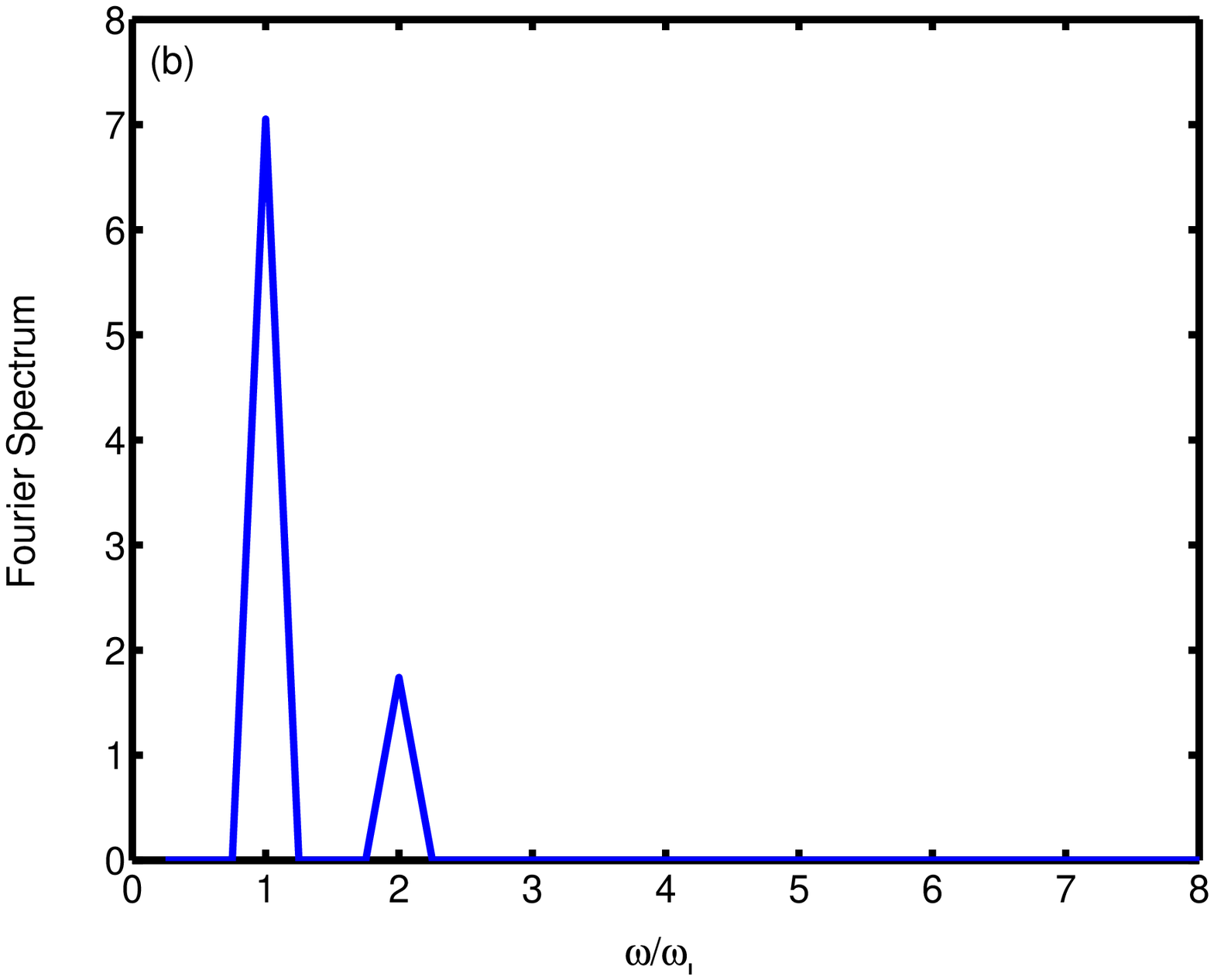,height=2.8cm,width=4cm,angle=0}}
\centerline{\psfig{figure=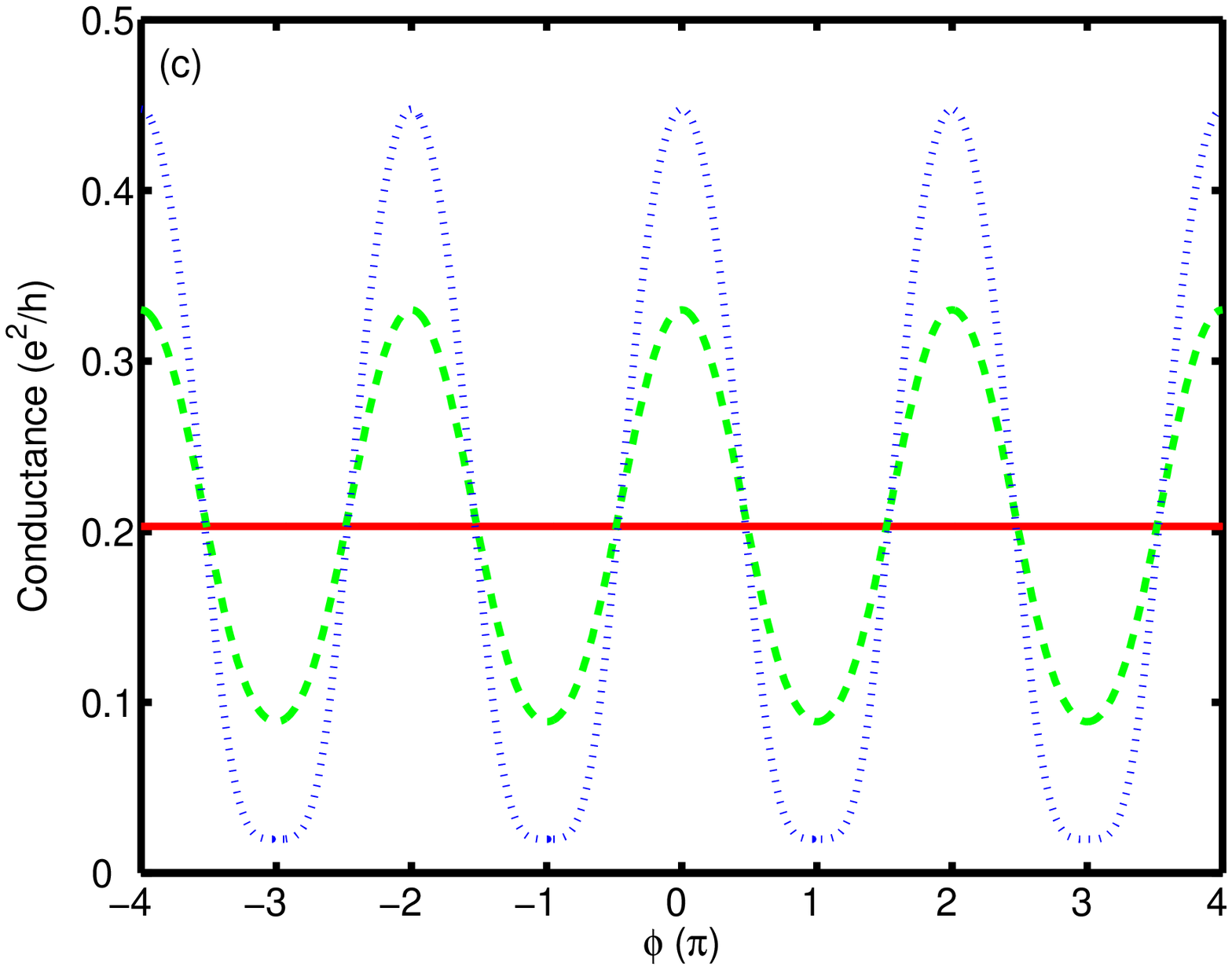,height=2.8cm,width=4cm,angle=0}
\psfig{figure=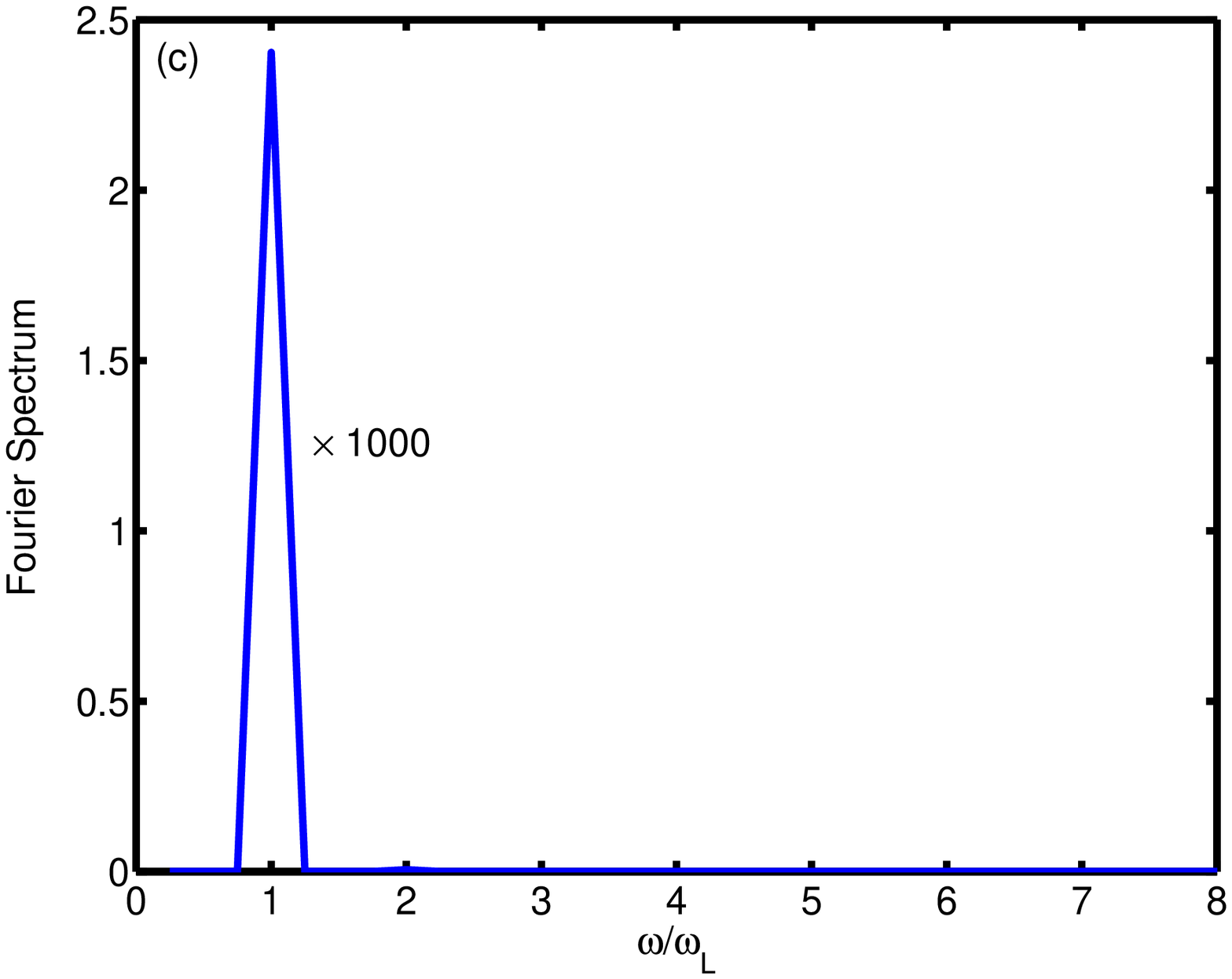,height=2.8cm,width=4cm,angle=0}}
\caption{Conductance versus the phase $\phi$ accumulated by the
precession of the local spin with various values of the spin-flip
coupling $\Gamma_{s}/\Gamma_{n}=0.0$ (red-solid line), 0.4
(green-dashed line), and 0.8 (blue-dotted line). The results shown
in the left panels (a) through (c) correspond to three different
values of the chemical potential: $\mu=0$, 0.4, and 0.8. Also
shown in the right panels is the Fourier spectrum for
$\Gamma_{s}/\Gamma_{n}=0.4$ with the chemical potential same as in
the left panels. Other parameter values: $\Gamma_{n}=0.1$,
$\theta=\pi/2$, and $\phi_0=0$.} \label{FIG:COND2}
\end{figure}

As we have shown, in the presence of magnetic field, the local
spin will precess with the Larmor frequency. The question is
whether this Larmor precession will manifest in the conductance of
electrons transported through this local spin, which we now
address below. To be relevant to the experimental situation, we
consider the electron tunneling in the off-resonance regime by
choosing the value of the chemical potential other than $\pm 1$.
In Fig.~\ref{FIG:COND2}, we plot the conductance as a function of
the phase $\phi$ accumulated by the precession of the local spin
for various values of the spin-flip couplings. Without loss of
generality, we have taken $\theta=\pi/2$ and $\phi_0=0$. The
results shown in the left panels (a)-(c) correspond to three
typical values of the chemical potential $\mu$=0, 0.4, and 0.8.
Also shown in the right panels is the Fourier spectrum for a fixed
value $\Gamma_{s}/\Gamma_{n}=0.4$ with the chemical potential same
as in the left panels. Several features are noteworthy: (i) The
conductance oscillation occurs only when the spin-flip couplings
is nonzero and its oscillation amplitude increases with the
spin-flip couplings. (ii) When $\mu=0$, the conductance exhibits
the periodicity of $\pi$ (see Fig.~\ref{FIG:COND2}(a)), which
corresponds to the oscillation of frequency $2\omega_{L}$. When
$\mu$ is nonzero, the conductance oscillates in phase with a basis
period of $2\pi$, that is, with frequency $\omega_{L}$ (see
Fig.~\ref{FIG:COND2}(b) and (c)). Generally, there still exists of
the $2\omega_{L}$ mode, the Fourier spectral weight of which
decreases with the deviation of $\mu$ from zero but is enhanced by
the spin-flip couplings. These modes can be seen more clearly from
the Fourier spectrum. Analytically, we find that the imaginary
part of the off-diagonal components of the retarded Green function
satisfy $\mbox{Im}G_{+-}^{r}(\phi+\pi)=\mbox{Im}G_{-+}^{r}(\phi)$.
Therefore, $\mbox{Im}[G_{+-}^{r}+G_{-+}^{r}]$ is a periodic
function of $\phi$ with a period of $\pi$. However, the
periodicity of $\mbox{Im}G_{++}^{r}$ and $\mbox{Im}G_{--}^{r}$
depends on the position of the chemical potential. With a little
algebra, one can obtain:
\begin{equation}
\mbox{Im}[G_{++}^{r}+G_{--}^{r}]=\frac{-4\omega(J\sin\theta
\Gamma_{s}\cos\phi+\omega\Gamma_{n})
+2\Gamma_{n}\tilde{\omega}^{2}} {\tilde{\omega}^{4}+
4(J\sin\theta\Gamma_{s}\cos\phi+\omega\Gamma_{n})^{2}}\;,
\label{EQ:GREEN-IMAGINARY}
\end{equation}
and
$\tilde{\omega}=[\omega^{2}-J^{2}-\Gamma_{n}^{2}+\Gamma_{s}^{2}]^{1/2}$.
Equation~(\ref{EQ:GREEN-IMAGINARY}) shows that the contribution to
the conductance from the spin-conserved couplings involves the
linear term and quadratic term in $\cos\phi$. When $\omega=0$, the
linear term vanishes, and the oscillation is determined solely by
the $\cos^{2}\phi$ term. This explains why  the conductance
oscillates with frequency $2\omega_{L}$ as $\mu=0$. Moreover,
since the contributions to the conductance from
$\mbox{Im}[G_{++}^{r}+G_{--}^{r}]$ and
$\mbox{Im}[G_{+-}^{r}+G_{-+}^{r}]$ are weighted by $\Gamma_{n}$
and $\Gamma_{s}$, respectively, one can expect that the spectral
weight of $2\omega_{L}$ mode in the conductance oscillation is
appreciable for a large ratio of $\Gamma_{s}/\Gamma_{n}$ even when
$\mu \neq 0$.

To conclude, we have studied the quantum transport through a local
impurity spin precessing in an external static magnetic field. We
have found that the spin-flip coupling between the conduction
electrons on the spin and those in the leads are crucial to the
appearance of the conductance oscillation. We have also predicted
that there exists the oscillation mode with frequency {\em twice
of the Larmor frequency}, the Fourier spectral weight of which can
be tuned by the position of the chemical potential and spin flip
coupling.

The following remarks are in order: We have assumed that the
dynamics of the local spin is controlled by the magnetic field
only and no decoherence mechanism is included, and have
concentrated on the forward action of the spin on the transport
properties of conduction electrons. In reality, the dynamics of
the local spin is also influenced by the backaction of the
transport currents due to its coupling to the conduction
electrons. Therefore, our results should be valid in the weak
measurement regime, where the spin relaxation time is sufficiently
long. In the opposite regime, the interactions of the spin with
its surroundings are so strong that the spin precession will die
out quickly and the spin is aligned with the magnetic field
$\mathbf{B}$.  One then can apply a small r.f. field perpendicular
to the static magnetic field. By solving the Bloch equation, one
can find the transverse components of the spin oscillate with the
r.f. frequency. This leads to an exchange interaction between a
driven spin and conduction electrons very similar to that given by
Eq.~(\ref{EQ:SPIN}). Therefore, the conductance can be evaluated
by following the same procedure and the conclusion about the
conductance oscillation remains. This setup is experimentally
accessible, the measurement from which will provide an additional
test of the proposed mechanism for the conductance oscillation.

{\bf Note Added}: While working on this paper, we became aware of
earlier work by Aronov and Lyanda-Geller~\cite{Aronov89}, where
the oscillation of average electron spin at the frequency of an
alternating electric field was shown.

{\bf Acknowledgments}: The authors thank Y. Manassen, Y. Meir, D.
Mozyrsky, and B. Spivak for helpful discussions. This work was
supported by the Department of Energy through the Los Alamos
National Laboratory.


\begin{thebibliography}{99}

\bibitem{Prinz98} G. Prinz, Science {\bf 282}, 1660 (1998).

\bibitem{Wolf01} S. A. Wolf {\em et al.}, Science {\bf 294}, 1488
(2001).

\bibitem{Kane98} B. E. Kane, Nature {\bf 393}, 133 (1998).

\bibitem{Loss98} D. Loss and D. P. DiVincenzo, Phys. Rev. A {\bf
57}, 120 (1998).

\bibitem{Farle98} M. Farle, Rep. Prog. Phys. {\bf 61}, 755 (1998).

\bibitem{Bruland98} K. J. Bruland {\em et al.}, Appl. Phys. Lett.
{\bf 73}, 3159 (1998).

\bibitem{Mana00} Y. Manassen, I. Mukhopadhyay, and N. Ramesh Rao,
Phys. Rev. B {\bf 61}, 16223 (2000); Y. Manassen, R. J. Hamers, J.
E. Demulth, and A. J. Castellano, Jr., Phys. Rev. Lett. {\bf 62},
2531 (1989); D. Shachal and Y. Manassen, Phys. Rev. B {\bf 46},
4795 (1992); Y. Manassen, J. Magentic Reson. {\bf 126}, 133
(1997).

\bibitem{Durkan02} C. Durkan and M. E.
Welland, Appl. Phys. Lett. {\bf 80}, 458 (2002).

\bibitem{Balats02} A. V. Balatsky and I. Martin, cond-mat/0112407.

\bibitem{Mozy01} D. Mozyrsky {\em et al.}, cond-mat/0201325.

\bibitem{Butt86} M. B\"{u}ttiker, Phys. Rev. Lett. {\bf 57}, 1761
(1986); R. Landauer, Philos. Mag. {\bf 21}, 863 (1970).

\bibitem{Note1} Here we do not address the difference between the
longitudinal ($T_1$) and transversal ($T_2$) relaxation time since
these times are defined for ensamble not single spins.

\bibitem{Meir92} Y. Meir and N. S. Wingreen, Phys.
Rev. Lett. {\bf 68}, 2512 (1992).

\bibitem{Aronov89} A. G. Aronov and Yu. B. Lyanda-Geller, Pis'ma
Zh. Eksp. Teor. Fiz. {\bf 50}, 398 (1989).

\end{thebibliography}
\end{document}